 \documentclass[10pt,preprint]{aastex63}
 \usepackage{graphicx,subfigure}

\def\a{\"a}
\def\o{\"o}
\def\u{\"u}
\def\A{\"A}
\def\O{\"O}
\def\U{\"U}

\def\gapprox{\lower.4ex\hbox{$\;\buildrel >\over{\scriptstyle\sim}\;$}}
\def\lapprox{\lower.4ex\hbox{$\;\buildrel <\over{\scriptstyle\sim}\;$}}

\shortauthors{Aschwanden and Araujo}
\shorttitle{SOC}

\begin{document}

\renewcommand{\topfraction}{0.95}
\renewcommand{\bottomfraction}{0.95}
\renewcommand{\textfraction}{0.05}
\renewcommand{\floatpagefraction}{0.95}
\renewcommand{\dbltopfraction}{0.95}
\renewcommand{\dblfloatpagefraction}{0.95}

\title{The Energy-Duration Relationship in Astrophysical 
Self-Organized Criticality Systems}
 
\author{Markus J. Aschwanden}
\affil{Lockheed Martin, Solar and Astrophysics Laboratory (LMSAL),
       Advanced Technology Center (ATC),
       A021S, Bldg.252, 3251 Hanover St.,
       Palo Alto, CA 94304, USA;
       e-mail: markus.josef.aschwanden@gmail.com}
\author[0000-0002-2106-4332]{Alexandre Araujo}
       \affiliation{Centro de Radio Astronomia e Astrofisica Mackenzie, 
       Universidade Presbiteriana Mackenzie, Rua da Consolacao, 
       930, Sao Paulo, SP, Brazil,
       e-mail: adesouza.astro@gmail.com}

\begin{abstract}
Scaling laws in astrophysical systems that involve
the energy, the geometry, and the spatio-temporal 
evolution, provide the theoretical framework for
physical models of energy dissipation processes.
A leading model is the standard fractal-diffusive
self-organized criticality (FD-SOC) model, which
is built on four fundamental assumptions: (i) the
dimensionality $d=3$, (ii) the fractal dimension
$D_V=d-1/2=2.5$, (iii) classical diffusion 
$L \propto T^{(1/2)}$, and (iv)
the proportionality of the dissipated energy to
the fractal volume $E \propto V$. Based on these 
assumptions, the FD-SOC model predicts a scaling
law of $T \propto E^k \propto E^{(4/5)} = E^{0.8}$.
On the observational side, we find empirical
scaling laws of $T \propto E^{0.81\pm0.03}$
by Peng et al.~(2023) and $T \propto E^{0.86\pm0.03}$
by Araujo \& Valio (2021) that are self-consistent
with the theoretical prediction of the FD-SOC model.
However, cases with a small time range 
$q_T = \log{(T_{max}/T_{min})} \lapprox 2$ have large
statistical uncertainties and systematic errors,
which produces smaller scaling law exponents 
($k \approx 0.3, ..., 0.6$) as a consequence.
The close correlation of the scaling exponent $k$
with the truncation bias $q_T$ implies that the
dispersion of k-values is an observational effect,
rather than a physical property.
\end{abstract}
 
\section{Introduction}

The concept of nonlinear systems governed by  
{\sl self-organized criticality (SOC)}
has been applied successfully to a large number
of astrophysical, geophysical, and other 
statistical datasets. There are two different
schools of thought that have been developed
in theoretical SOC models: the cellular automaton
approach, which we call the Bak-Tang-Wiesenfeld
(BTW) model (Bak et al.~1987), and the 
fractal-diffusive (FD-SOC) model (Aschwanden 2014). 
The former model is faciliated with
numerical (next-neighbor interaction) simulations,
while the latter is expressed in terms of an 
analytical (physical scaling law) approach
(see textbooks Aschwanden 2011, 2025 and
references therein).

Regarding quantitative modeling and predictions
there are three SOC parameters of interest:
the power law slope $\alpha_x$ of (differential and
cumulative) size (occurrence frequency)
distribution functions, the power law slope
$\alpha_{\rm WT}$ of waiting time distributions,
and physical scaling laws $T \propto E^k$.
In this Letter we focus on the third relationship,
which entails the duration $T$ of a SOC avalanche,
the energy $E$ or flux $F$, and the scaling
exponent $k$. The test presented here consists of 
analyzing 11 published relationships $T \propto E^k$ by
comparing the observed values with the theoretial
predictions, which represents a new test
of the FD-SOC model.
The robust correlation of the scaling exponent $k$
with the duration range $q_T = \log{(T_{max}/T_{min})}$ 
implies a truncation bias that is an observational effect,
rather than a physical property.

The contents of this Letter is a brief
description of the observations (Section 2),
a brief description of the relevant
theoretical FD-SOC model (Section 3), 
data analysis and results (Section 4),
and the conclusions (Section 5).

\section{Observations}

The statistical studies detected a correlation between 
energy and duration time for solar flares, stellar 
superflares, and prompt emission of GRBs with photons 
of keV to MeV, i.e., $T \propto E^k$ 
(Hou et al.~2013;      
Maehara et al.~2015;   
Namekata et al.~2017;  
Tu and Wang 2018;      
Peng et al.~2023,      
Araujo and Valio 2021, 
and Cai et al.~2024).  
We summarize these published datasets in chronological order.

\subsection{Hou et al. (2013)}

The dataset of Hou et al.~(2013) is based on gamma-ray bursts
(GRB) observed with the Fermi and Swift spacecraft, where
$T$ is the GRB duration and $E$ is the energy. 
The fitted scaling laws yield 
$E \propto T^{0.62\pm0.23}$,
$E \propto T^{0.56\pm0.03}$, and 
$E \propto T^{0.99\pm0.15}$.
However, since the axes are displayed and fitted in reverse
order, $E \propto T^{k,inv}$, rather than with the standard 
representation used here, i.e., $T \propto E^k$, they are 
incompatible with the other parameters of this study 
and thus are discarded in the following.

\subsection{Maehara et al.(2015)} 

The dataset of Maehara et al.~(2015) shows a correlation
between the duration $T$ of superflares as a function of
the bolometric energy $E$ (Fig.~1). The filled squares
and small crosses indicate superflares on G-type main
sequence stars detected from short- and long-cadence data
with the Kepler spacecraft. The flare duration is measured
from e-folding decay times. The dotted line indicates
the linear regression for the data of superflares from
short-cadence data, yielding a power law fit of 
$T \propto E^{0.39\pm0.03}$ (Fig.~1). There is an
effective threshold at $T \gapprox 3$ min visible in
Fig.~(1), which could cause a truncation bias in
the scaling law exponent $k$, especially since the
fitted time range is only $q_T \approx 1.5$ decades.
Furthermore, the flare threshold was defined by 3 times
the top 1\% of the size distribution, which contributes 
to a size-dependent truncation bias.

\subsection{Namekata et al.~(2017)}

A correlation between the flare durations $T$ and
flare energies $E$ are shown in Fig.~(2), taken
from the dataset of Namekata et al.~(2017).
The filled squares are solar {\sl white-light flares
(WLFs)}, and the gray crosses are areas on M-type star 
GJ 1243. Note that the durations 
are defined as that from the beginning to the end of 
flares, and the energies are converted into that in the
Kepler passband by assuming a 10,000 K blackbody.
The WLFs were observed with HMI/SDO. The power law
slope is $T \propto E^{0.38\pm0.06}$ (Fig.~2) 
(Namekata et al.~2017).

\subsection{Tu and Wang (2018)}

A correlation between the isotropic energy and duration
of {\sl gamma-ray bursts (GRBs)} has been discovered in Swift
data, $T \propto E^{0.34\pm0.03}$ (Fig.~3). After comparing with 
solar flares from RHESSI and stellar superflares from
the Kepler data, similar correlations were found with
$T \propto E^{0.33\pm0.001}$ (Fig.~4) and 
$T \propto E^{0.39\pm0.025}$ (Fig.~5).

\subsection{Peng et al. 2023}

In a survey observed with the Fermi-Large Area Telescope (LAT), 
data from solar gamma-ray flares and from 3C 454.3 GRBs
were analyzed, such as the duration $T$, the isotropic energy $E$,
the peak luminosity $L_P$, and the following correlations were found
in four datasets (Peng et al.~2023): 
$T \propto E^{0.81\pm0.08}$ for the Sun (Fig.~6),
$T \propto E^{0.61\pm0.01}$ for 3C 454.3 (Fig.~7),
$T \propto E^{0.38\pm0.08}$ (Fig.~8), and 
$T \propto E^{0.31\pm0.06}$ (Fig.~9).

\subsection{Arajo and Valio 2021}

Detecting starspots in a study of the activity of the star Kepler-411
a correlation between the starspot area and their durations
was found in Araujo and Valio (2021), $T \propto E^{0.86\pm0.03}$ 
(Fig.~10). The energy is assumed to be proportional to the
starspot area.

\subsection{Cai et al.~(2024)}

Analyzing flare observations from the {Heliospheric Magnetic Imager
(HMI)} onboard the {Solar Dynamics Observatory (SDO)}, a set of 
{\sl white-light flares (WLF)} were detected (Cai et al.~2024), 
with a correlation of $T \propto E^{0.25}$ (Fig.~11).

\section{Theoretical Model}

The {\sl fractal-diffusive self-organized criticality (FD-SOC)}
model (Aschwanden 2014, 2015, 2022, 2025)
is based on four fundamental assumptions:
(i) The fractality,
(ii) the scale-freeness,
(iii) the flux-volume proportionality of incoherent processes, and
(iv) classical diffusion.
Based on these four assumptions, the FD-SOC model predicts
power law functions for the size distributions of SOC parameters,
as well as the power law slopes for each distribution.
The FD-SOC model makes quantitative predictions as a function of
the Euclidean space dimension $d$, i.e.,
$d=1$ for curvi-linear structures,
$d=2$ for area-like geometries, and
$d=3$ for voluminous structures.
In the following we use only the Euclidean dimension of $d=3$,
which applies to most structures in the observed 3-D world,
which we call the {\sl standard SOC} model.
In practice, a mean value of $D_V=2.5$ corresponds to SOC
avalanches that are neither fully space-filling ($d=3$)
nor purely filamentary ($d=2$), but occupy the volume in a
highly intermittent fractal manner ($d\approx 2.5$).

The spatial inhomogeneity of a SOC avalanche is expressed
in terms of the fractal dimension $D_d$.
Each fractal domain has a maximum fractal dimension of $D_d=d$,
a minimum value of $D_d=(d-1)$, which implies a mean value of
$D_V=d-1/2=2.5$,
\begin{equation}
        D_V={(D_{\rm V,max} + D_{\rm V,min}) \over 2} = d-{1 \over 2}  \ .
\end{equation}
The fractal geometry essentially corrects for the ubiquitous
spatial inhomogeneity that is present in most
cellular automaton topologies.
In the following we denote $D_3$ also as $D_V$.
The fractal volume $V$ is defined by the standard
(Hausdorff) fractal dimension $D_V$ in 3-D and the length scale
$L$ (Mandelbrot 1977),
\begin{equation}
        V \propto L^{D_V} \ .
\end{equation}
which yields the scaling $V \propto L^{(5/2)}$ for $d=3$ space.
The flux $F$ or energy $E$ is assumed to be proportional to the 
avalanche volume $V$ for incoherent growth ($\gamma=1$),
\begin{equation}
        E \propto F \propto V = \left( L^{D_V} \right)^{\gamma} \ .
\end{equation}
while coherent growth can be parameterized with 
$\gamma \gapprox 2$, a parameter that is not used in the 
standard FD-SOC model.
 
The spatio-temporal evolution is approximated with the assumption
of (classical) diffusive transport ($\beta=1)$,
\begin{equation}
        L \propto T^{\beta/2} = T^{1/2} \ ,
\end{equation}
with the classical diffusion coefficient $\beta=1$.

Combining Eqs.(2) and (3) allows us to eliminate the
variable $L$, while Eqs.~(2) and (4) yield the scaling law
between the energy $(E)$ (or flux $F$) and duration $T$,
\begin{equation}
	T \propto E^k  
	\propto E^{(2/D_V)} \propto E^{4/5} = E^{0.8} \ .
\end{equation}
Note that the scaling law exponent $k=0.8$ depends only on
the fractal dimension $D_V$ and the diffusion coefficient
$\beta=1$,
\begin{equation}
	k = {2 \over \beta D_V} ={2 \over D_V} = 0.8 \ .
\end{equation}
The energy-duration relationship $T \propto
E^k$ is equally valid for the energy $E$, the fluence, 
the flux $F$, or the fractal volume $V$, 
\begin{equation}
	T \propto E^{0.8} \propto F^{0.8} \propto V^{0.8} \ .
\end{equation}
as long as the proportionality assumption for incoherent 
emission holds as a suitable approximation. 
In summary, the standard FD-SOC model assumes 
($d=3, \gamma=1, \beta=1$).
The scaling law exponent $k$ is universal within the assumptions
of the standard FD-SOC model (i.e., dimension $d=3$, classical
diffusion $\beta=1$, and incoherent emission $\gamma=1$). 
There are a 
number of emission mechanisms in high-temperature plasmas
that obey this assumption, such as thermal bremsstrahlung,
incoherent gyroemission, or incoherent gyrosynchrotron 
emission, (e.g., see textbook by Benz 1993).
 
\section{Data Analysis and Results}

\subsection{Data}

We analyze now the published 11 datasets, which were selected
by the availability of information on the energy-duration 
correlations observed in white-light flares (WLF) and
gamma-ray bursts (GRB), summarized in Table 1 and
plotted in Figs.~(1)-(11). Table 1 contains the dataset number
(column 1), the scaling exponent $k$ in the scaling law
$T \propto E^k$ (column 2), the logarithmic time range
$q_T=\log{(T_{max}/T_{min})}$ (column 3), the physical units 
of the SOC parameters (energy, fluence, flux) (column 4),
the astrophysical phenomenon (WLF, GRB) (column 5),
the observing instruments or spacecraft (Kepler,
HMI/SDO, RHESSI, Kepler, Fermi, Hinnode/SOT) (column 6), and
the original literature references (column 7). 

\subsection{Time Duration Ranges}

The time duration range in the 11 datasets,
$q_T=\log(T_{max}/T_{min})$, varies
from small ranges $q_T \approx 1.0 = \log(10)$ 
to the largest ranges of $q_T \approx 3.0 = 
\log(1000)$. Since power law fits
in SOC statistics are most accurate for large
ranges $q_T \gapprox 2 = \log(100$), the scaling
exponents $k$ are also more accurate and reliable
for large duration ranges ($q_T = \log(100) \gapprox 2$).
We are plotting the scaling exponents $k$ as a
function of the logarithmic time range $q_T$
in Fig.~(12). The scaling exponents $k$ are
clearly correlated with the time ranges $q_T$
(see diamonds and solid line in Fig.~12). 
Datasets with large time
ranges $q_T \ge 2$ show a scaling exponent of
$k \approx 0.8$, which matches our theoretial
predictions of the FD-SOC model, 
while the observations reveal 
$k=0.81\pm0.08$, Peng et al.~(2023); and
$k=0.86\pm0.03$, Araujo and Valio (2021).
Datasets with small time
ranges $q_T \lapprox 2$ show a much smaller 
scaling exponent of $k \approx 0.3, ... , 0.4$, 
which cannot be reconciled with the FD-SOC model. 
The apparent mismatch of the theoretical
prediction can naturally be explained by the
large scatter of the correlated SOC parameters 
$T$ vs. $E$, which is most likely due
the truncation bias in small-number statistics, 
caused by inconsistent event detection 
techniques, incomplete statistical sampling, 
finite system size effects, and
random noise, as demonstrated in numerous
SOC studies that contain uncertainty estimates
of power law fitting (Aschwanden 2025).
Consequently, an explanation
of this shortcoming should be investigated
in terms of these systematic errors,
rather than in terms of statistical uncertainties
(which always suggest unrealistic small values,
see column 2 in Table 1).

\subsection{Truncation Bias}

The truncation bias can be modeled empirically 
with a linear function between the minimum
time range at $k(q_T = 0.5)=0.0$ and the
maximum time range at $k(q_T \approx 2.0)=0.8$
predicted by the FD-SOC model (Fig.~12, dashed line),
\begin{equation}
	k(q_T) = 
	\left\{
	\begin{array}{lll}
	\ 	& 0.4\ (q_T-0.5) & {\rm for}\ 0.5 \le q_T \le 2.5 \\
	\	& 0.8            & {\rm for}\         q_T \ge 2.5 
	\end{array}
	\right.
\end{equation}
Fig.~12 demonstrates that this empirical linear
function (dashed curve in Fig.~12) closely matches 
the observed values of the scaling exponent $k$ (curve
with diamonds in Fig.~12). An observational test of
this prediction is finding new datasets with sufficiently
large time ranges ($q_T \gapprox 2.0$), which should then
exhibit power law coefficients of $k \approx 0.8$,
while smaller datasets should exhibit values
$k(q_T)$ according to the empirical model defined
in Eq.~(8).

\subsection{Alternative Physical Models}

The empirical truncation bias explains
the large deviations from the theoretical FD-SOC model
and suggests that alternative physical scaling
laws are not required, such as SOC models in terms
of Alfv\'enic travel times (Tu and Wang 2018; 
Peng et al.~2023; Cai et al.~2024).
One has also to take into account that the definition
of durations may depend on the assumed physical model, 
such as conductive loss, radiative cooling, 
hydrogen recombination, or magnetic reconnection. 
Nonthermal emissions (in hard X-rays) generally 
produce shorter pulses than thermal
emission (in soft X-rays), which can be characterized
by the time integral of the hard X-ray flux (according
to the Neupert effect known in solar flares;
Neupert and Zarro 1993). 

\section{Conclusions} 

We investigate physical scaling laws operating
in SOC avalanches, which should provide a deeper
understanding of the observed correlation between time
duration $T$ and energy $E$, i.e., $T \propto E^k$,
in SOC avalanches. The theoretical prediction
of the scaling exponent is $k=2/D_V=4/5=0.8$,
based on the fractal dimension of $D_V=5/2=2.5$
and the diffusion coefficient $\beta=1$.
This theoretical prediction is matched in the
observations in 2 cases:
$k=0.81\pm0.08$, Peng et al.~2023; and
$k=0.86\pm0.03$, Araujo and Valio 2021).

The other 9 analyzed cases reveal much
flatter scaling exponents, in the order of
$k \approx 0.3, ..., 0.4$, which are most
likely due to various sources of systematic
errors, such as the truncation bias due to
small-number statistics, inconsistent event 
detection techniques, incomplete statistical 
sampling, and finite system size effects. 
The argument of a truncation
bias due to small-number statistics is further
corroborated by the dependence of the scaling
exponent $k(q_T)$ on the duration range $q_T$ 
(Fig.~12), which spans from the possible
minimum value $k(q_T=0.5)=0$ to the 
maximum value $k(q_T{\ge 2.5})=0.8$  
in an approximately linear relationship
(Fig.~12). Thus this study is capable to
predict the energy-duration relationship
as a function of the sampled time ranges,
$k(q_T)$, which represents the first attempt
to explain the energy-duration relationship
in self-organized criticality systems.
The main contribution of the paper is its
role as a critical synthesis, providing a
unified interpretation of  previously
reported, apparently inconsistent
energy-duration scaling exponents.
The close correlation of the scaling exponent $k$
with the truncation bias $q_T$ implies that the
dispersion of k-values is an observational effect,
rather than a physical property.

\acknowledgements
This work was stimulated by the organizers of a
workshop on “Mechanisms for extreme event generation”
(MEEG) at the Lorentz Center at Snellius, Leiden,
The Netherlands, July 8-12, 2019, organized by Drs.
Norma Bock Crosby, Bertrand Groslambert, Alexander
Milovanov, Jens Juul Rasmussen, and Didier Sornette.
The author acknowledges the hospitality and partial
support of two previous workshops on “Self-Organized
Criticality and Turbulence” at the International
Space Science Institute (ISSI) at Bern, Switzerland,
during October 15-19, 2012, and September 16-20, 2013,
as well as constructive and stimulating discussions
with Sandra Chapman, Paul Charbonneau, Aaron Clauset,
Norma Crosby, Michaila Dimitropoulou, Manolis
Georgoulis, Stefan Hergarten, Henrik Jeldtoft Jensen,
James McAteer, Shin Mineshige, Laura Morales, Mark
Newman, Naoto Nishizuka, Gunnar Pruessner, John Rundle,
Carolus Schrijver, Surja Sharma, Antoine Strugarek,
Vadim Uritsky, and Nick Watkins. This work was partially
supported by NASA contracts NNX11A099G “Self-organized
criticality in solar physics”, NNG04EA00C of the SDO/AIA
instrument, and NNG09FA40C of the IRIS instrument.
The authors acknowledge partial funding from Brazilian 
agencies CNPq (grants 150817/2022-3 and 172886/2023-6).

\section*{References}
\def\ref#1{\par\noindent\hangindent1cm {#1}}
 
\ref{Araujo, A. and Valio, A.
	2021, ApJL 922:L23 (7pp)}
	{\sl Kepler-411 star activity between starspots
	and superflares}
\ref{Aschwanden, M.J.
        2011, Self-organized criticality in astrophysics. The
        statistics of nonlinear processes in the universe,
	Springer/PRAXIS: Berlin}  
\ref{Aschwanden, M.J. 
        2014, ApJ 782, 54, }
        {\sl A macroscopic description of self-organized systems and
        astrophysical applications}
\ref{Aschwanden, M.J.
 	2022, ApJ 934:33 (27pp),} 
 	{\sl The fractality and size distributions of 
	astrophysical self-organized criticality systems}
\ref{Aschwanden, M.J.
        2025, Power Laws in Astrophysics. 
	Self-Organzed Criticality Systems, 
	Cambridge University Press: Cambridge.}
\ref{Bak, P., Tang, C., and Wiesenfeld, K.
 	1987, PhRvL 59/4, 381-384}
 	{\sl Self-organized criticality: 
	An explanation of 1/f noise}
\ref{Benz, A.O.,
	1993, Plasma Astrphysics. Kinetic processes
	in solar and stellar Coronae, Kluwer
	Academic publishers: Dordrecht: Netherlands.}
\ref{Cai, Y., Hou, Y., Li, T., Liu, J.
	2024, ApJ 975:69}
	{\sl Statisgtics of solar white-light flares.
	I. optimization and application of identification
	methods}
\ref{Hou, S.J., Liu, T., Lin, D.B., Wu, X.F, and Lu, J.F.
	2012, Proc. IAU Symp. No. 290, 223,
	C.M. Zhang, Belloni, T. and Mendez, S.N. Zhang, (eds.)}
	{\sl Is there a relation between duration and $E_{iso}$
	in gamma-ray bursts?}
\ref{Maehara, H., Shibayama, T., and Notsu, Y.
        2015, Earth, Planet and Space 67, id.59, 10pp.}
        {\sl Statistical properties of superflares on 
	solar-type stars based on 1-min cadence data}
\ref{Mandelbrot, B.B. 1977,
        Fractals: form, chance, and dimension, Translation of
        Les objects fractals, W.H. Freeman, San Francisco.}
\ref{Namekata, K., Sakaue, K., Watenaba, K., Asai, A., et al.
	2017, ApJ 851:91.}
	{\sl Statistical studies of solar white-light flares
	and comparisons with superflares on solar-type stars}
\ref{Dennis, B.R. and Zarro,D.M.
 	1993, Sol.Phys. 146, 177-190}, 
 	{\sl The Neupert effect: what can it tell us about 
	the impulsive and gradual phases of solar flares}
\ref{Peng, F.K., Hou, S.J., Zhang, H.M., Xue, R., and
	Shu, X.W. 2023, MNRAS 520, 5974-5981}
	{\sl Similar properties between gamma-ray
	emission of 3C 454.3 and solar GeVflares}
\ref{Tu, Z.I., and Wang, F.Y.
	2018, ApJ 869, L23}
	{\sl The correlation between isotropic energy and
	duration of gamma-ray bursts}
\clearpage

\begin{table}
\begin{center}
\caption{Scaling exponent $k$ of correlation $T \propto E^{k}$
and time range $\log(T_{max}/T_{min})$ are tabulated  
in columns 2 and 3. Astronomical phenomena
include gamma-ray bursts (GRB) and White-light flares (WLF).
The (high-lighted) datasets 6, 7, and 10 exhibit a time range of
$\log{(T_{max}/T_{min}}\ge 2.0$ and have a power law slope that
is self-consistent to the theoretical FD-SOC model.}
\medskip                
\begin{tabular}{cllllll}
\hline
Dataset & Scaling & Time & Physical & Astrophysical & Instrument & Reference \\
        & exponent& range & unit    & phenomenon    & Spacecraft &           \\
        & k       & [decades] &     &               &            &           \\
\hline
\hline
1       & 0.39$\pm$0.03  & 1.5 & Energy & WLF       & Kepler     & Maehara et al.(2015)\\
2       & 0.38$\pm$0.06  & 1.5 & Energy & WLF       & Kepler, Hinode/SOT & Namekata et al.(2017)\\
3       & 0.34$\pm$0.03  & 1.5 & Energy & GRB       & Swift      & Tu and Wang (2018)\\ 
4       & 0.33$\pm$0.001 & 1.5 & Counts & GRB       & RHESSI     & Tu and Wang (2018)\\ 
5       & 0.39$\pm$0.03  & 1.5 & Energy & GRB       & Kepler     & Tu and Wang (2018)\\ 
{\bf 6} & {\bf 0.81}$\pm$0.08  & {\bf 3.0} & Fluence& GRB        & Fermi      & Peng et al.(2023)\\
{\bf 7} & {\bf 0.61}$\pm$0.01  & {\bf 2.0} & Energy & GRB        & Fermi      & Peng et al.(2023)\\
8       & 0.38$\pm$0.08  & 1.5 & Fluence& GRB       & Fermi      & Peng et al.(2023)\\
9       & 0.31$\pm$0.08  & 1.0 & Energy & GRB       & Fermi      & Peng et al.(2023)\\
{\bf 10}& {\bf 0.86}$\pm$0.03  & {\bf 2.0} & Energy & WLF        & Kepler     & Araujo and Valio (2021)\\
11      & 0.25           & 1.0 & Energy & WLF       & HMI/SDO    & Cai et al.(2024)\\
\hline
        & 0.80           & $\ge 2.0$ & Energy & Universal &      & FD-SOC theory\\
\hline
\end{tabular}
\end{center}
\end{table}

\clearpage

\begin{figure}
\begin{center}
\includegraphics[width=0.5\textwidth,angle=-90]{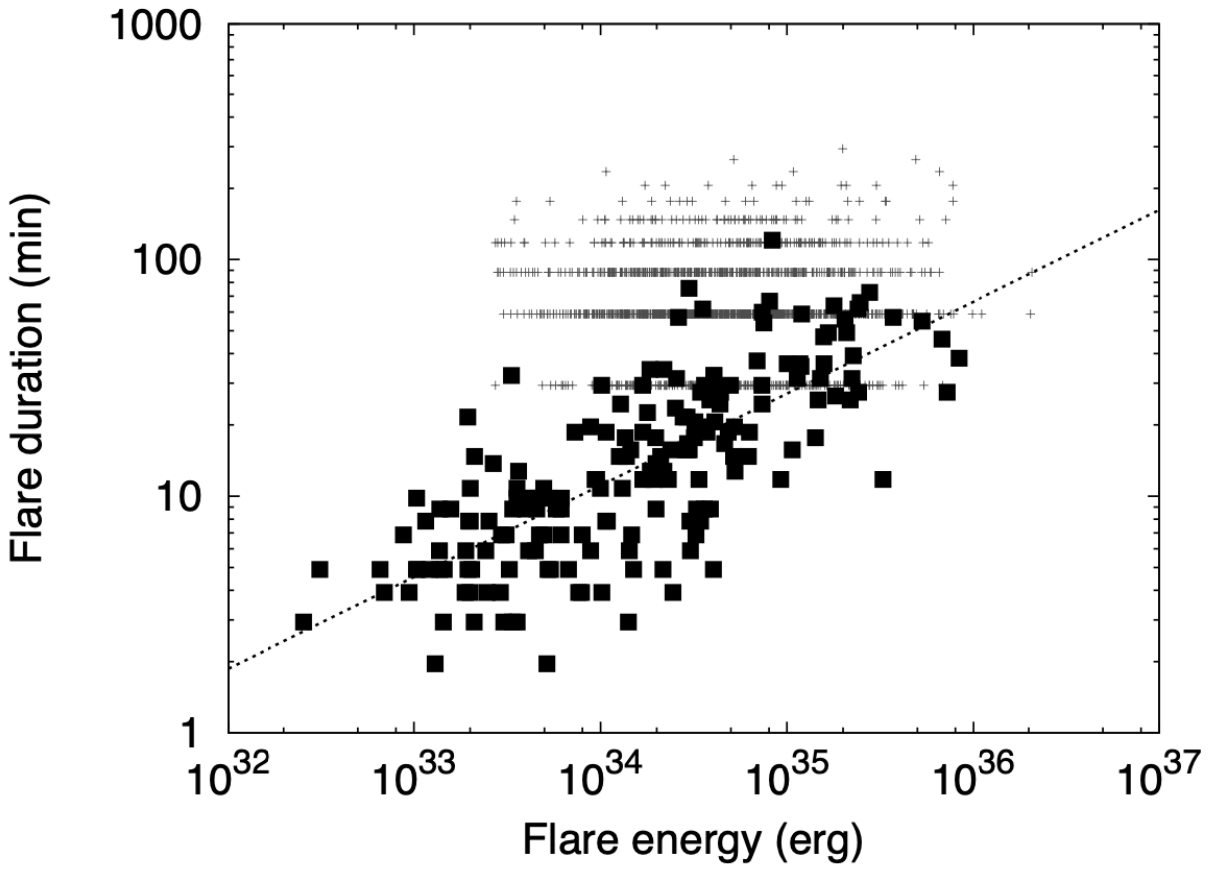}
\caption{Duration vs. Energy (Maehara et al.~2015).}
\end{center}

\begin{center}
\includegraphics[width=0.5\textwidth,angle=-90]{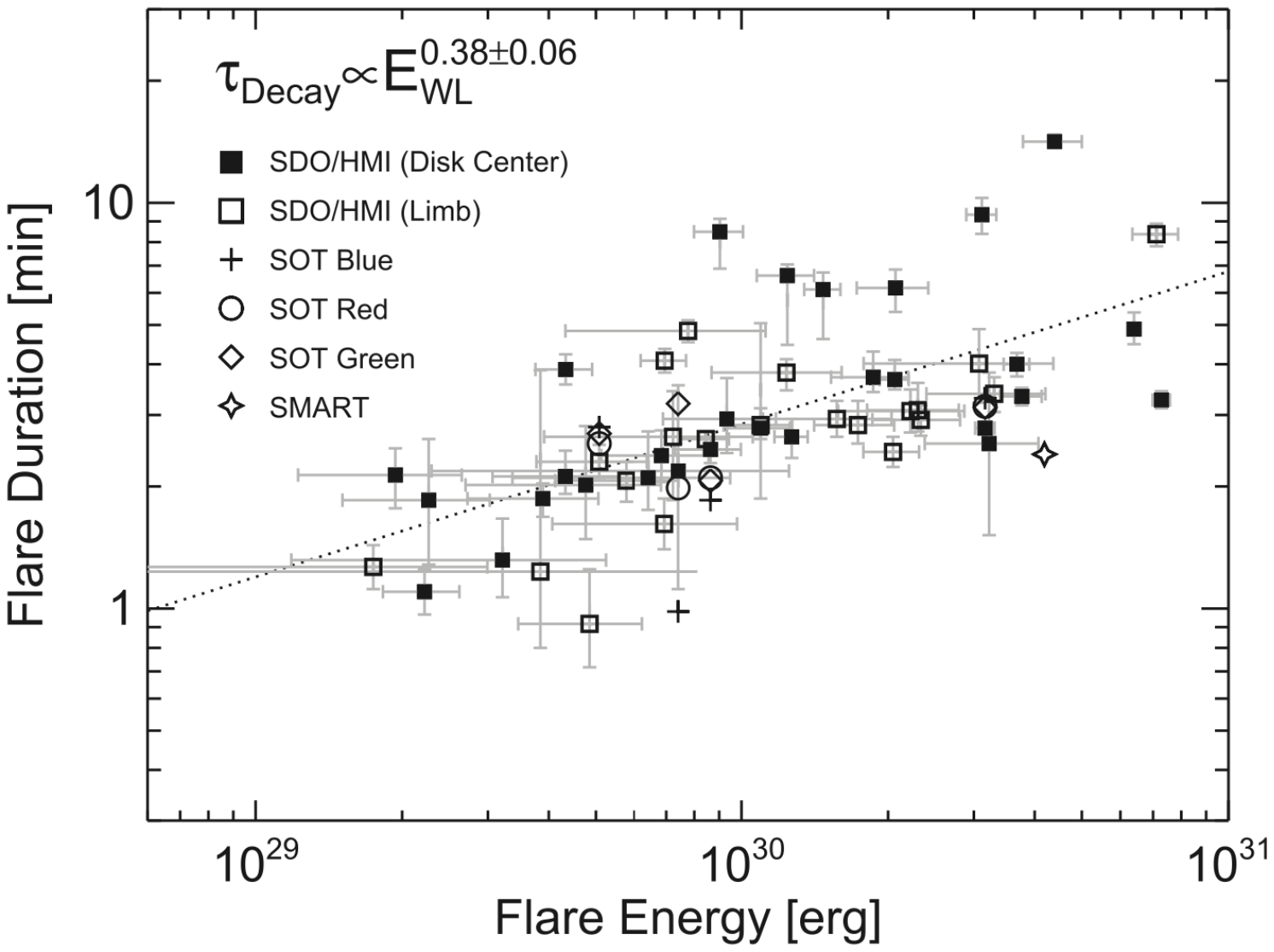}
\caption{Duration vs. Energy (Namekata et al.~2017)}
\end{center}
\end{figure}
 
\begin{figure}
\begin{center}
\includegraphics[width=0.5\textwidth,angle=-90]{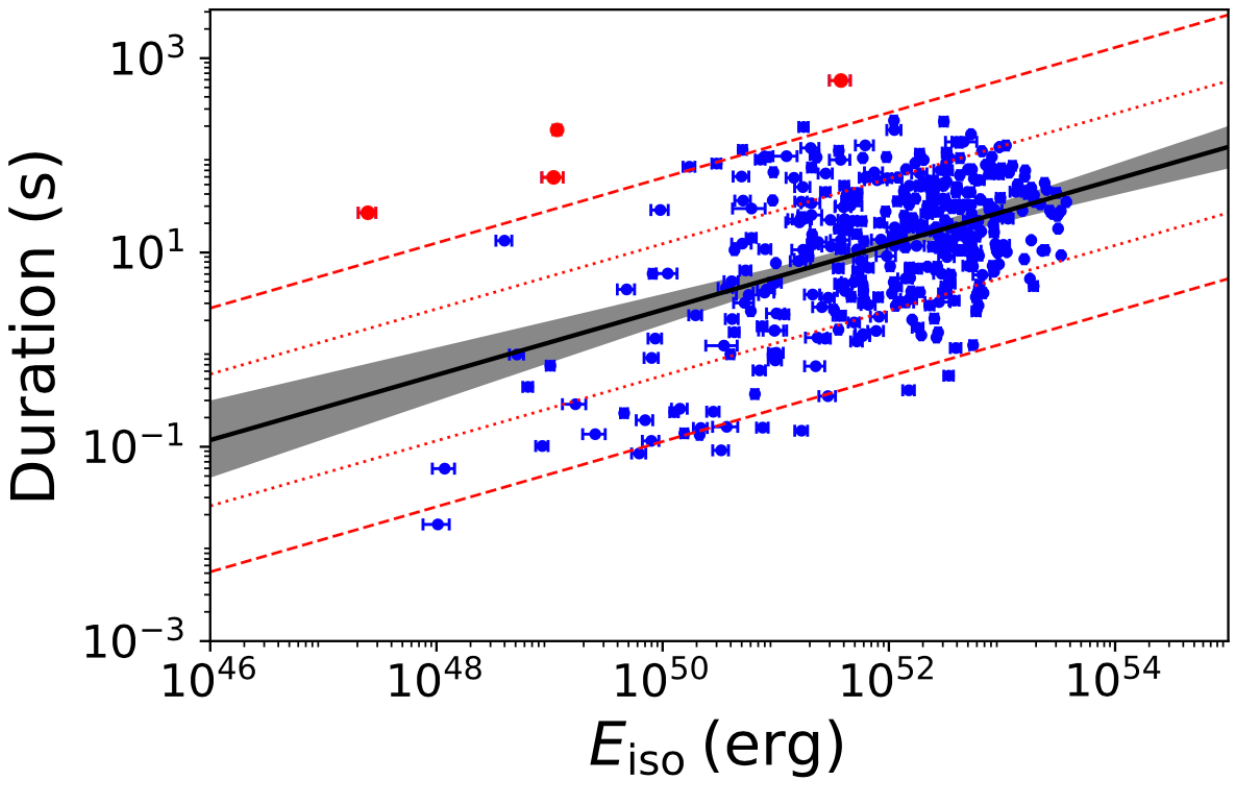}
\caption{Duration vs. Energy (Tu and Wang 2018)}
\end{center}

\begin{center}
\includegraphics[width=0.5\textwidth]{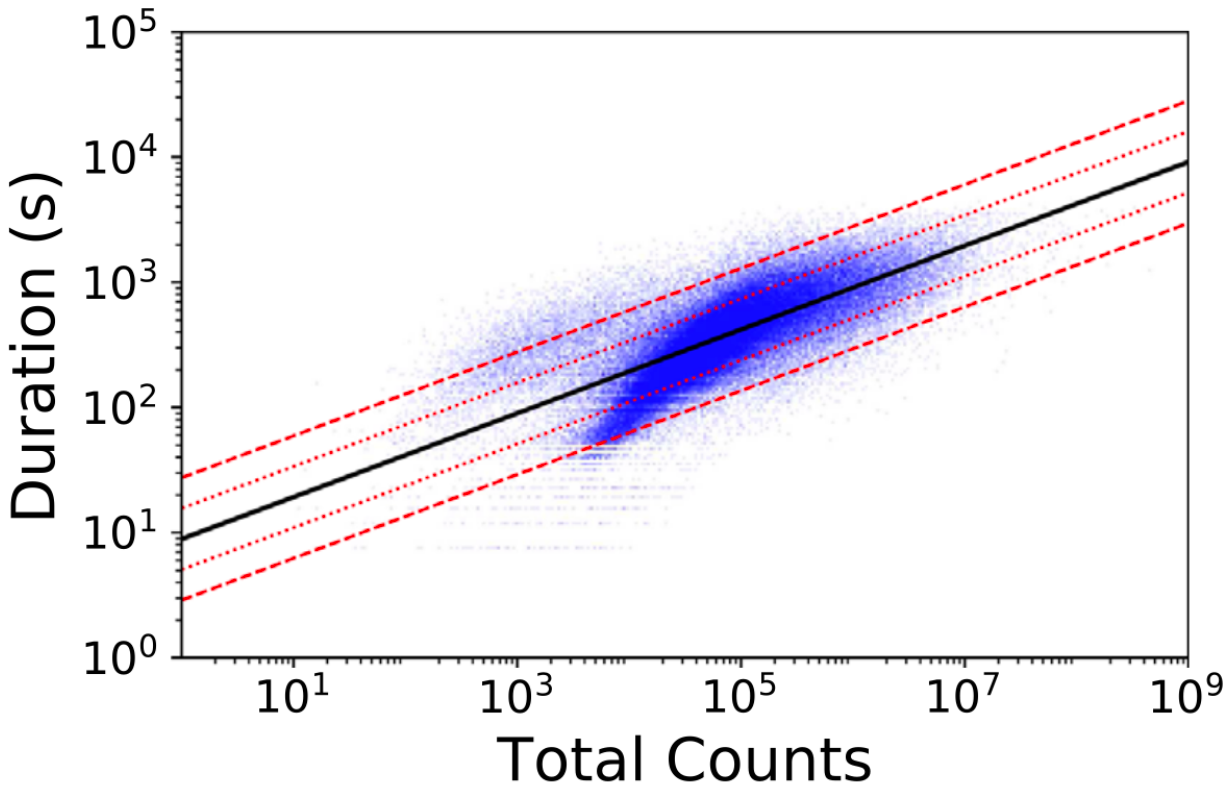}
\caption{Duration vs. Flux (Tu and Wang 2018)}
\end{center}
\end{figure}

\begin{figure}
\begin{center}
\includegraphics[width=0.5\textwidth,angle=-90]{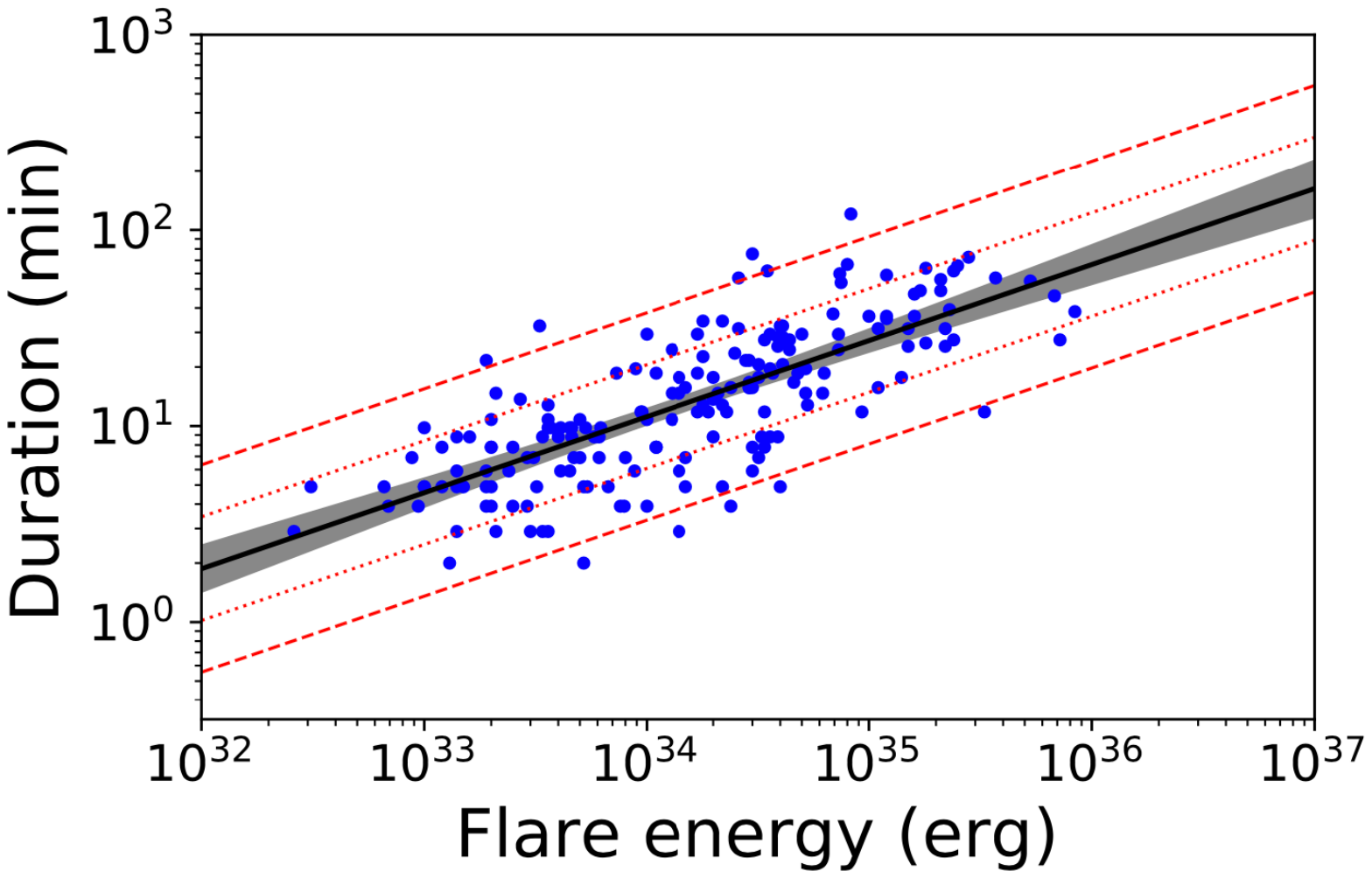}
\caption{Duration vs. Energy (Tu and Wang 2018)}
\end{center}

\begin{center}
\includegraphics[width=0.5\textwidth,angle=-90]{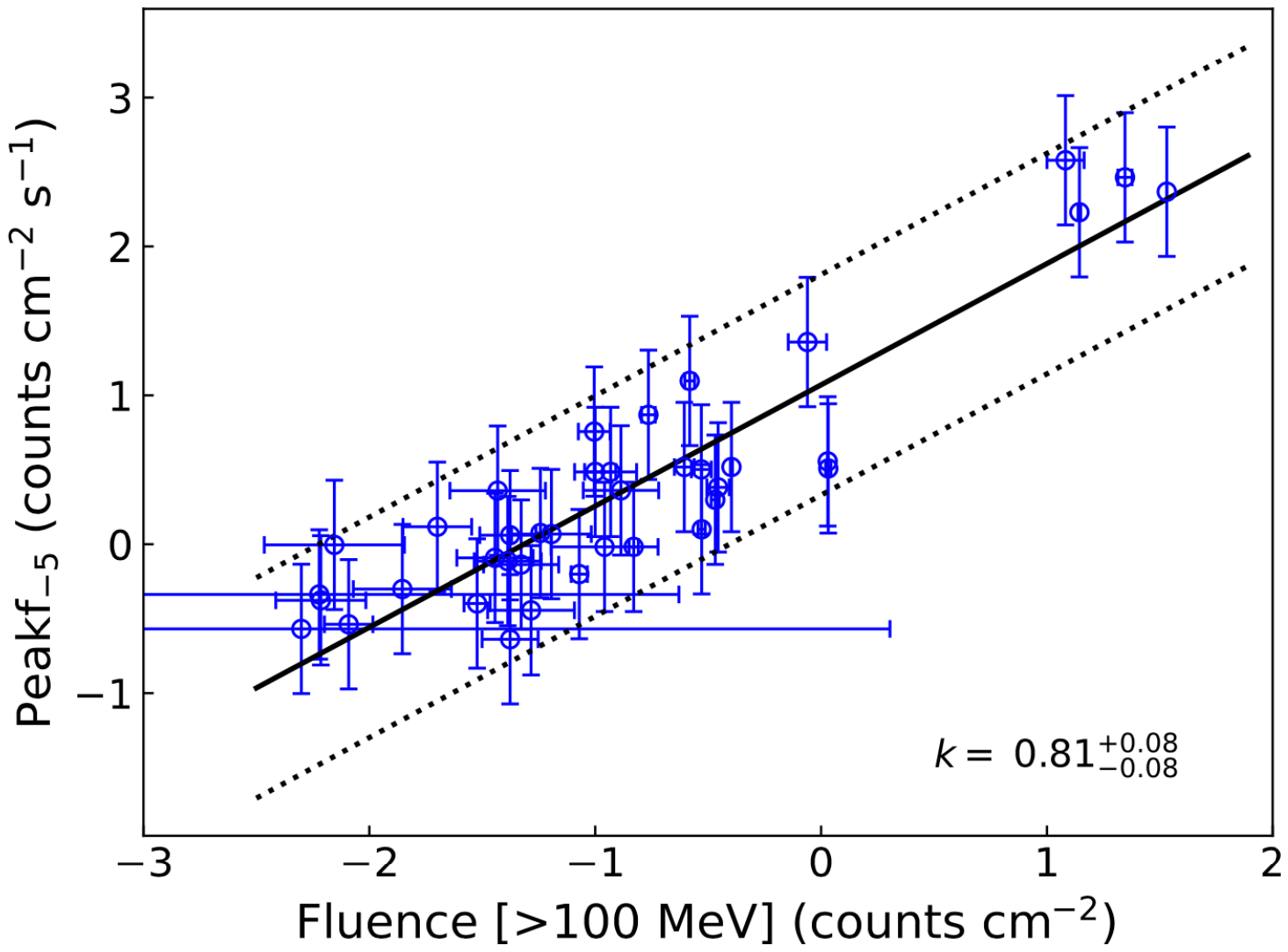}
\caption{Duration vs. Fluence (Peng et al.~2023)}
\end{center}
\end{figure}

\begin{figure}
\begin{center}
\includegraphics[width=0.5\textwidth,angle=-90]{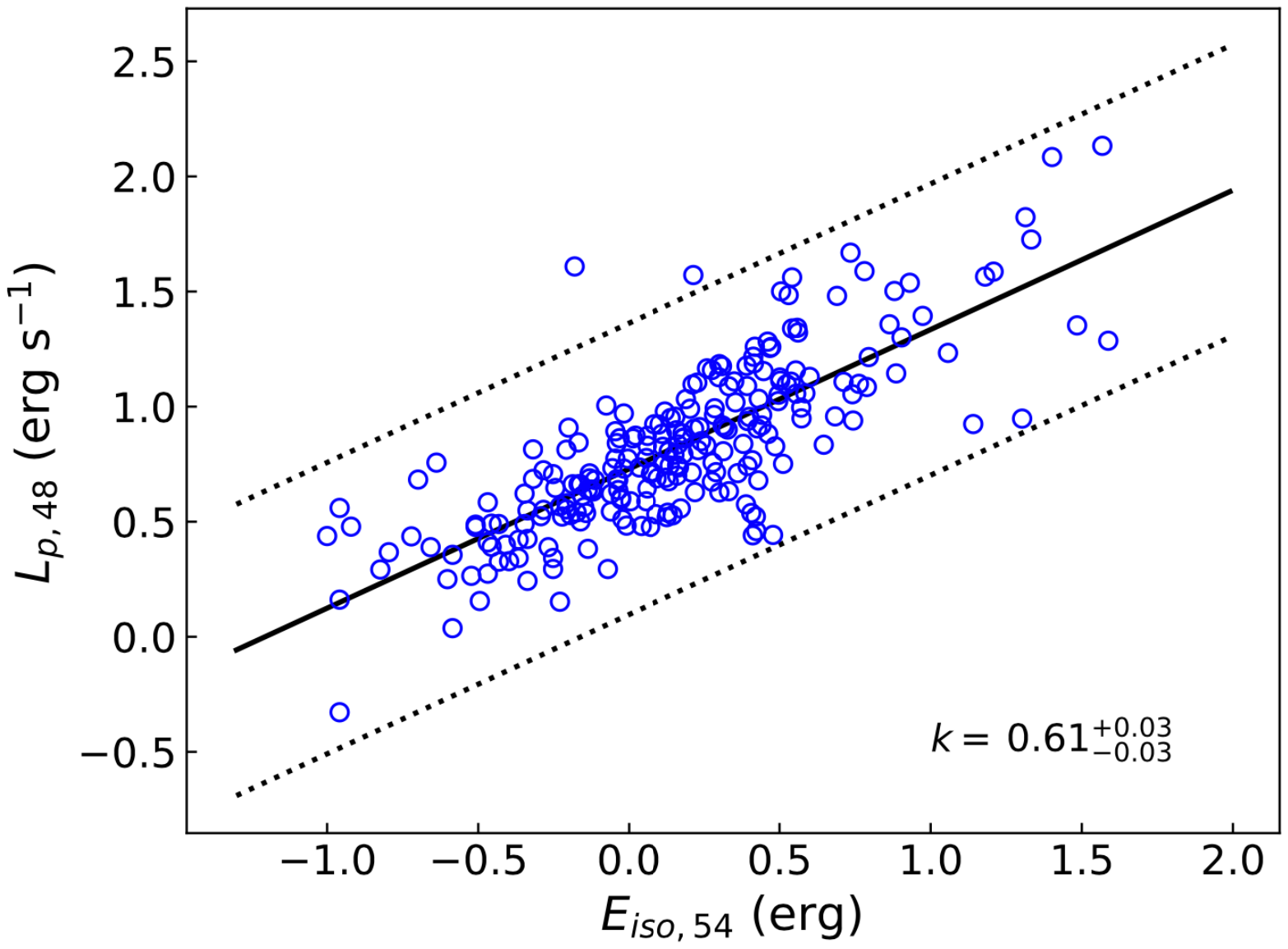}
\caption{Duration vs. Energy (Peng et al.~2023)}
\end{center}

\begin{center}
\includegraphics[width=0.5\textwidth,angle=-90]{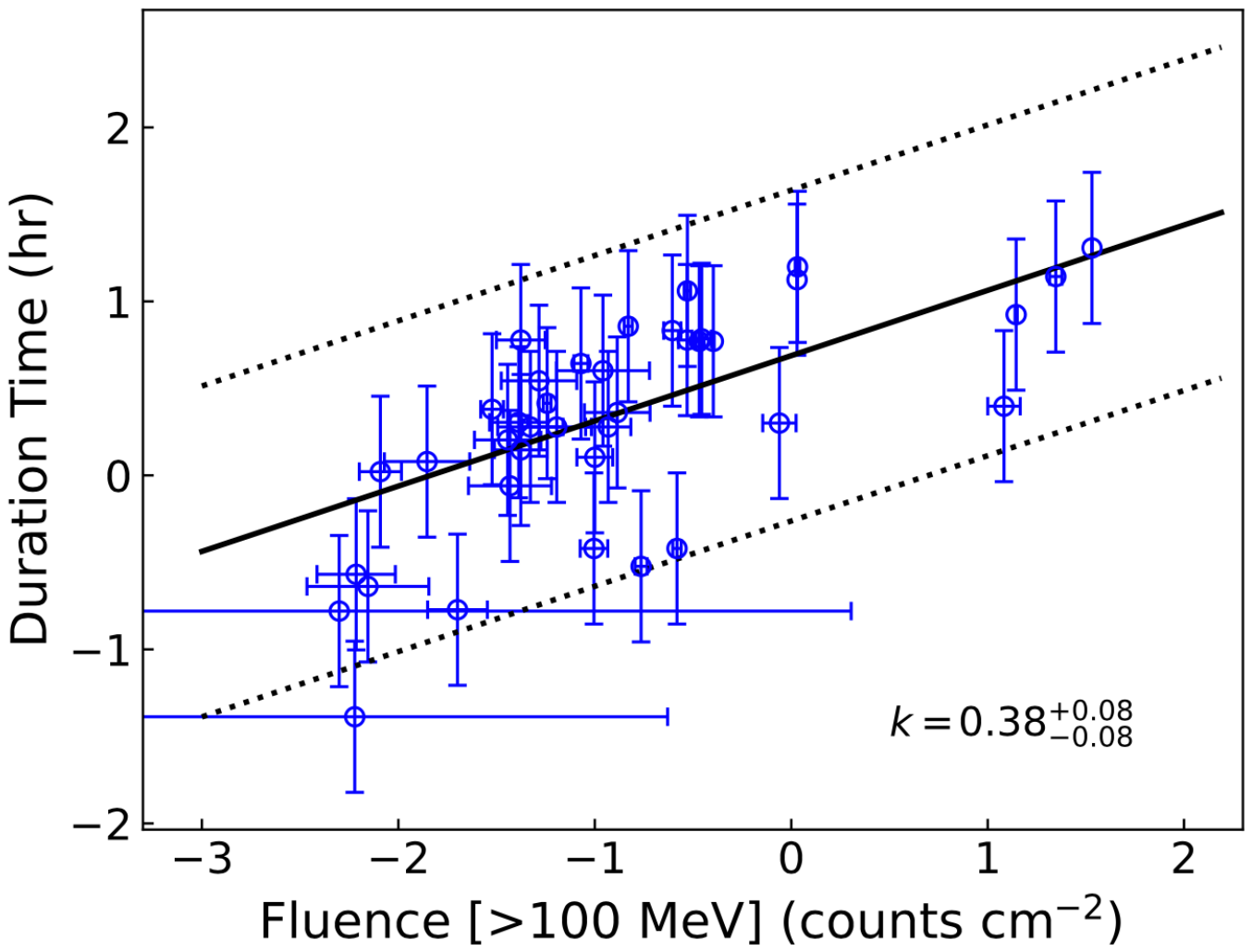}
\caption{Duration vs. Fluence (Peng et al.~2023)}
\end{center}
\end{figure}

\begin{figure}
\begin{center}
\includegraphics[width=0.5\textwidth,angle=-90]{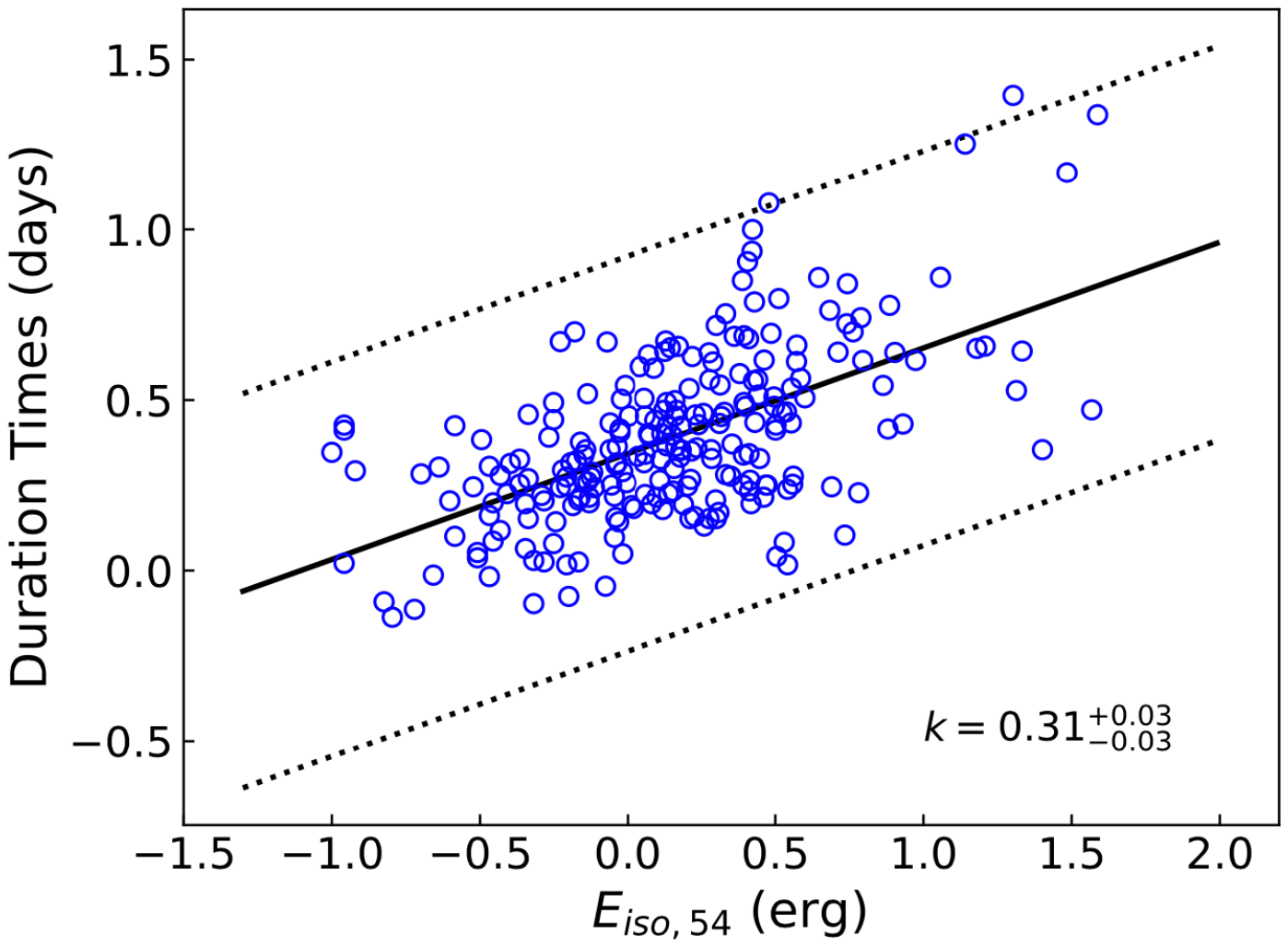}
\caption{Duration vs. Energy (Peng et al.~2023)}
\end{center}
\end{figure}

\begin{figure}
\begin{center}
\includegraphics[width=0.5\textwidth,angle=-90]{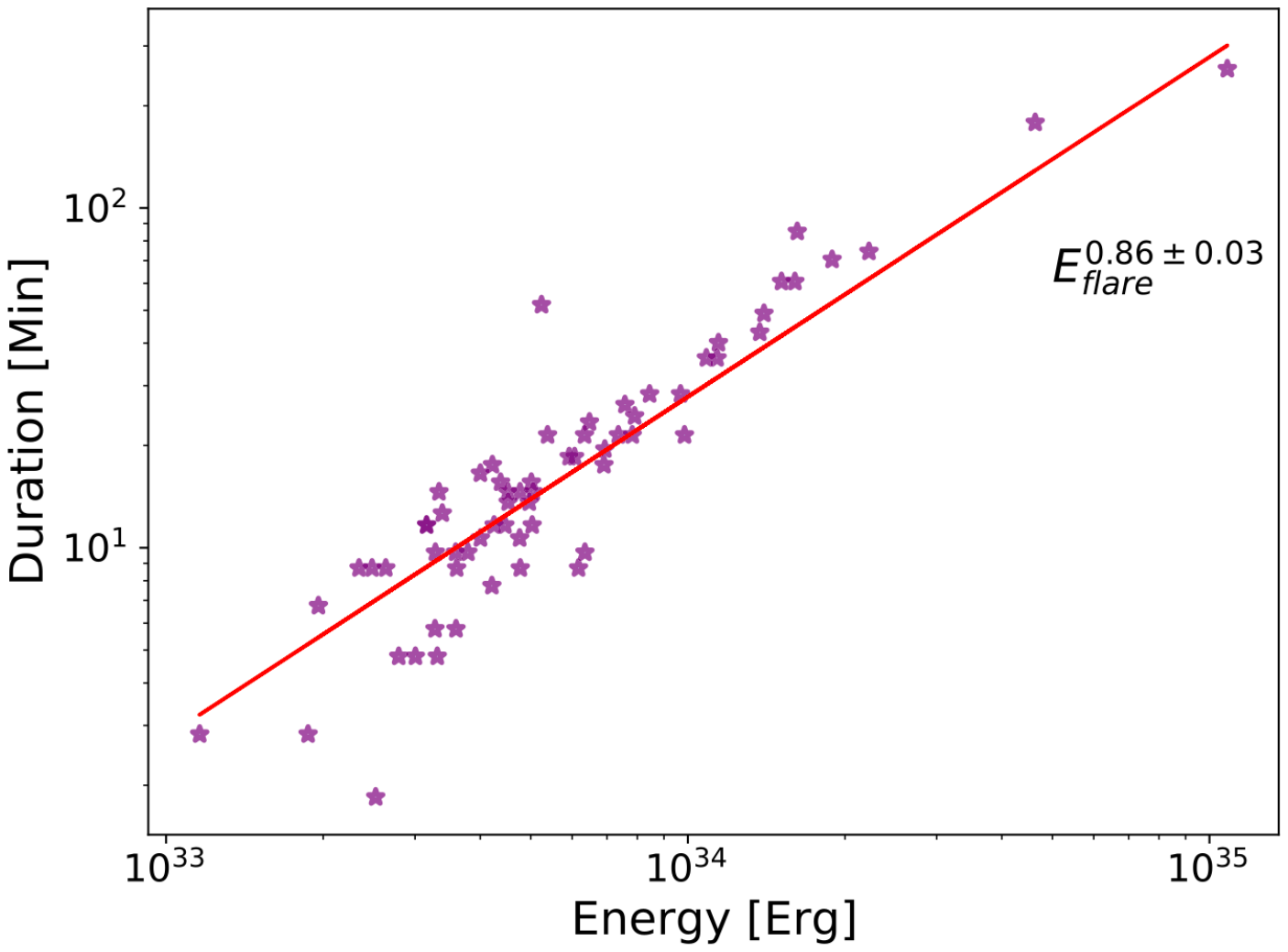}
\caption{Duration vs. Energy (Araujo and Valio 2021)}
\end{center}
\end{figure}

\clearpage
\begin{figure}
\begin{center}
\includegraphics[width=0.5\textwidth,angle=-90]{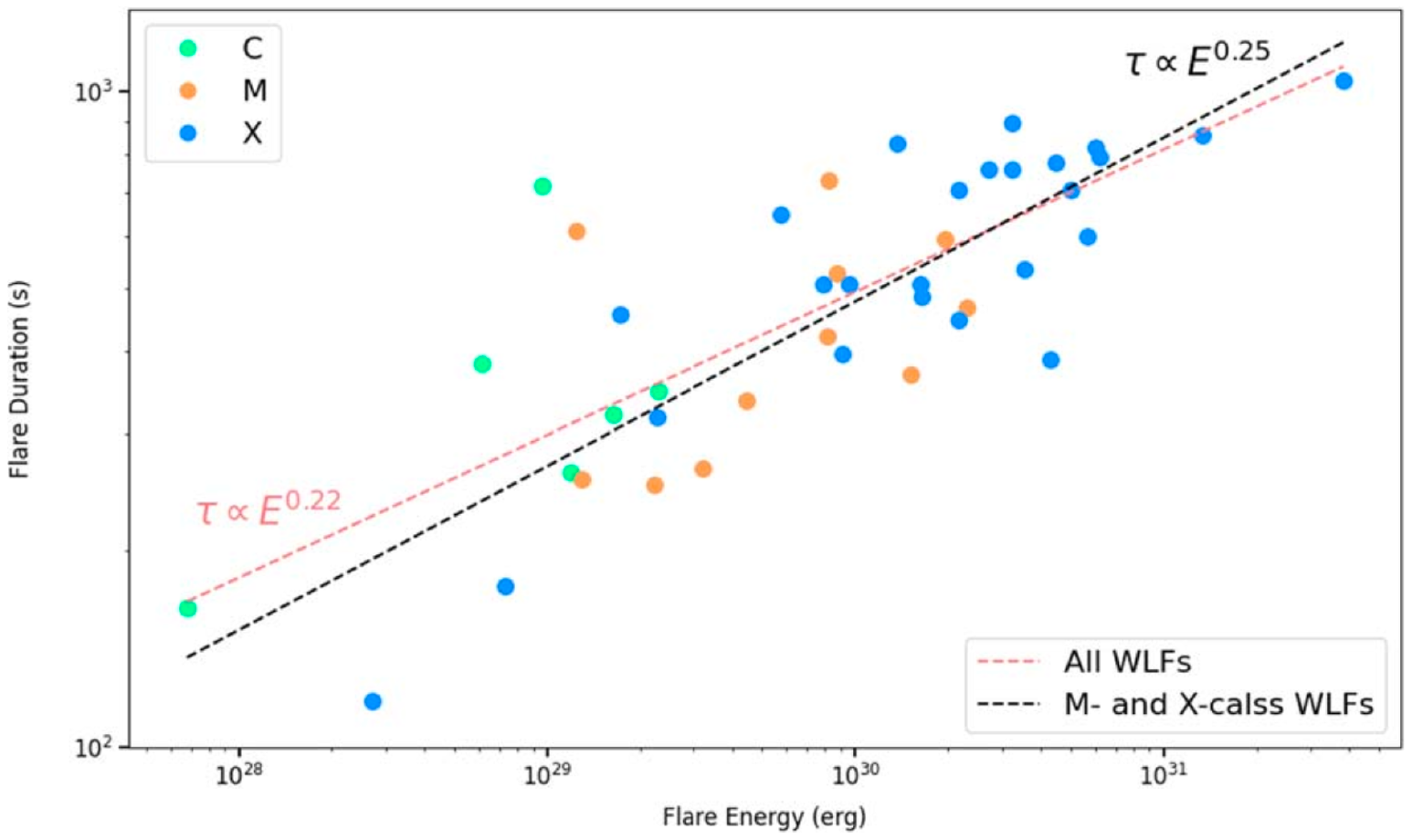}
\caption{Duration vs. Energy (Cai et al.~2024).}
\end{center}
\end{figure}

\begin{figure}
\begin{center}
\includegraphics[width=0.8\textwidth]{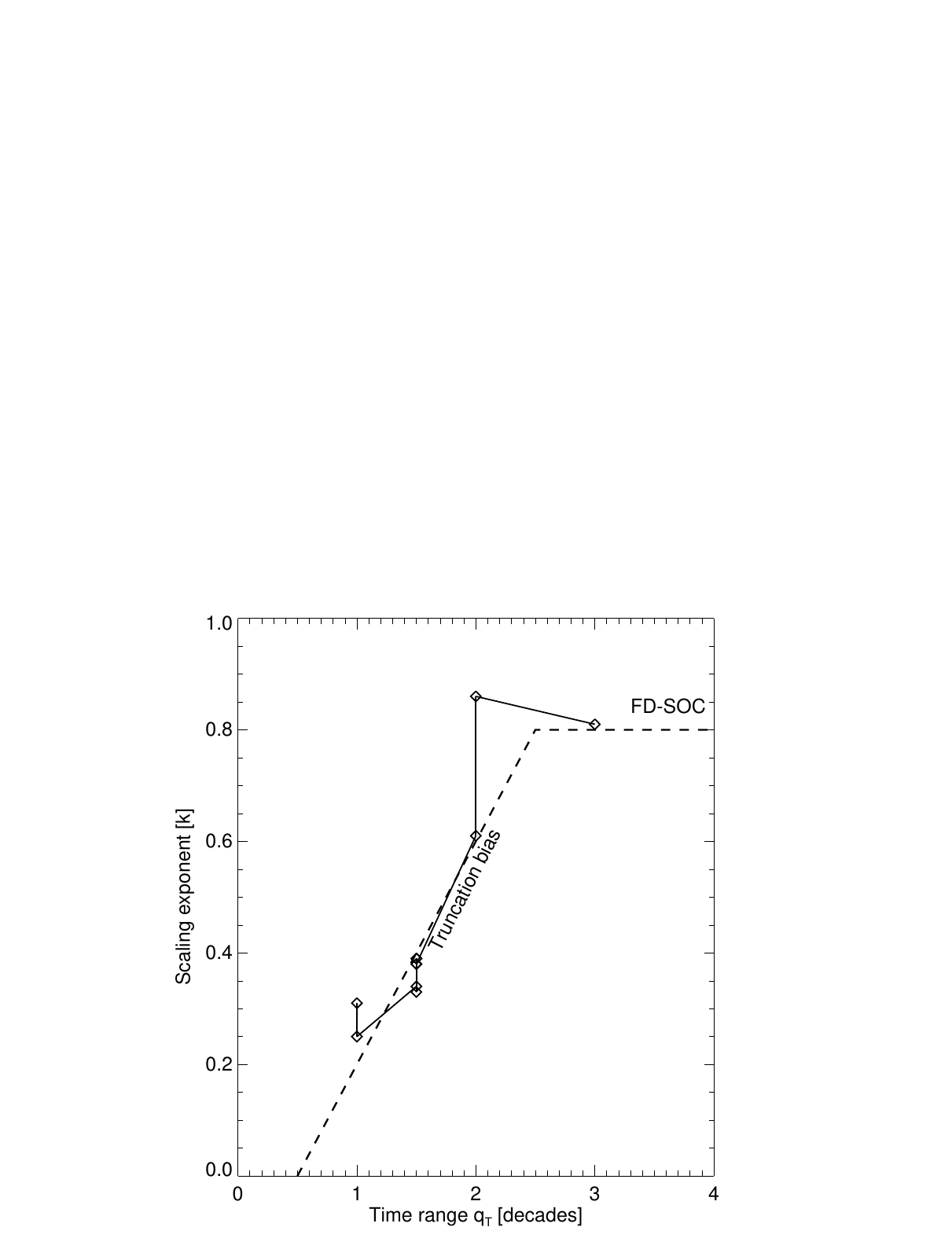}
\caption{Scaling exponent $k$ as a function of the
time range $q_T = \log(T_{max}/T_{min})$ in the scaling law
$T \propto E^k$. The dashed line indicates an empirical
truncation bias for small time ranges ($q_T \lapprox 2.0$),
and the prediction $q_T=0.8$ of the FD-SOC model for large 
time ranges $q_T \gapprox 2.0$.}
\end{center}
\end{figure}

\end{document}